\documentclass[aps, twocolumn, prb, floatfix, superscriptaddress,longbibliography]{revtex4-2}
\usepackage{amsmath,amssymb,amsthm}
\usepackage{xcolor}
\usepackage{graphicx}
\usepackage{appendix}

\newcommand{\tr}

\newcommand{\be}{\begin{equation}}
\newcommand{\ee}{\end{equation}}
\def\ba{\begin{aligned}}
\def\ea{\end{aligned}}
\newcommand{\bea}{\begin{eqnarray}}
\newcommand{\eea}{\end{eqnarray}}
\def\bes{\begin{subequations}}
\def\ees{\end{subequations}}
\def\bal{\begin{align}}
\def\eal{\end{align}}

\usepackage{soul}
\usepackage{ulem} 

\usepackage{amsmath,amssymb,bm}
\usepackage{enumerate}
\usepackage{braket}        

\makeatletter
\let\selectlanguage\@gobble

\makeatother

\begin{document}

\title{Nonlinearity-induced Topological Phase Transitions in One-dimensional Systems with Chiral Symmetry: A Real-Space Treatment}

\author{Liu Bi Tao}%
\affiliation{Department of Physics, National University of Singapore, Singapore 117542, Singapore}
\author{Shilpi Roy}
\affiliation{Centre for Quantum Technologies, National University of Singapore, Singapore 117543, Singapore}
\author{Jiangbin Gong}%
\email{phygj@nus.edu.sg}
\affiliation{Department of Physics, National University of Singapore, Singapore 117542, Singapore}
\affiliation{Centre for Quantum Technologies, National University of Singapore, Singapore 117543, Singapore}

\date{\today}

\begin{abstract}
Nonlinearity can induce topological phase transitions in periodic systems. It is hence of considerable interest to examine if such transitions persist in the absence of translational symmetry. We address this question directly in real space, using the periodic Su–Schrieffer–Heeger (SSH) chain as a reference for understanding the effect of quasiperiodicity. How  nonlinearity drives a chain towards a topological  phase transition is found to depend strongly on the strength of the quasiperiodic potential. As the nonlinearity strength increases,  we find that weak quasiperiodicity  supports a transition from a trivial to a topological phase, whereas the intermediate regime does not have a topological phase transition.  Under strong quasiperiodicity, pronounced boundary localization can develop well before a nonlinearity-induced topological transition.  Our results show how nonlinearity may interplay with a quasiperiodic potential in inducing topological phase transitions and that boundary localization of nonlinear eigenstates alone may not be taken as a clear signature of topological edge states.
\end{abstract}

\maketitle


\section{\label{sec:level1}Introduction}

Topological phases of matter have attracted tremendous interest due to their distinctive properties, most notably their robustness against perturbations and disorder \cite{hasanKaneTopologicalInsulators2010,QiZhang2011,Schnyder2008,BernevigHughes2013}. While the majority of existing studies focus on linear topological band theory, recent works have begun to explore how nonlinearity modifies and enriches topological physics \cite{soneNonlinearityinducedTopologicalPhase2024,hadadSelfinducedTopologicalTransitions2016,tuloupNonlinearityInducedTopological2020,tuloupTopologicalCharacteristicsGap2022,bomantaraNonlinearDiracCones2017,Ezawa_2022,Alharthy_Bomantara_2026,smirnovaNonlinearTopologicalPhotonics2020}. Nonlinear effects can give rise to a range of novel phenomena in the context of topological physics, including self-induced topological states~\cite{lumerSelfLocalizedStates2013, leykamEdgeSolitonsNonlinearPhotonic2016}, topological solitons~\cite{guoWeaklyNonlinearTopological2020, maTopologicalEdgeSolitons2021}, nonlinear Dirac cones with a quantized nonlinear Berry phase~\cite{bomantaraNonlinearDiracCones2017}, the breakdown of quantized Thouless pumping~\cite{Tuloup_Bomantara_Gong_2023}, and multistability in topological modes~\cite{baiArbitrarilyConfigurableNonlinear2024} which may coexist with or compete against conventional notions of topological protection. Moreover, nonlinear systems offer enhanced tunability through parameters such as interaction strength and drive amplitude~\cite{baiArbitrarilyConfigurableNonlinear2024, dobrykhNonlinearControlElectromagnetic2018}, making them promising platforms for realizing controllable and adaptive topological phases.

Nonlinearity-induced topological phenomena have been studied in both momentum space and real space. For periodic systems treated in real space,  previous works have reported nonlinearity-induced topological transitions and the accompanying changes from bulk delocalized states to edge-localized states \cite{hadadSelfinducedTopologicalTransitions2016,tuloupNonlinearityInducedTopological2020}, as well as nonlinear edge states and solitons, including their evolution, stability, and topological character \cite{maTopologicalEdgeSolitons2021,jezequelNonlinearEdgeModes2022}. More recently, real-space approaches have enabled the topology of nonlinear systems to be characterized without relying on translational invariance \cite{wong2023probing}.

Quasiperiodic systems lack translational invariance while retaining long-range spatial order, providing a distinct setting between periodic and randomly disordered systems \cite{Aubry1980,DominguezCastro2019}. Their localization properties have been extensively studied across a wide range of experimental platforms, including ultracold atoms in incommensurate optical lattices, photonic lattices, acoustic systems, and electrical circuits \cite{Roati2008,Lahini2009,Modugno2009,apigo2019observation}. The interplay between quasiperiodicity and topology has also attracted considerable interest. In particular, quasiperiodic modulation can drive linear SSH-type systems from trivial to topological phases \cite{lu2022exact,tang2022topological}.

Nonlinear topological phenomena have also been studied in quasiperiodic lattices \cite{rosa2023amplitude,li2026topological,schneider2024topological}. Increasing nonlinearity can modify or destroy pre-existing edge localization and can also generate localized nonlinear states from bulk states. However, in these studies topology was either inherited from the linear system or not independently characterized in the nonlinear regime. 
This raises the question of whether nonlinearity itself can induce topology in an initially trivial quasiperiodic system, and how quasiperiodicity modifies real-space features through such a transition.

In this work, we address this question by studying nonlinearity induced topological transitions directly in real space -- with the real space approach being necessary due to the lack of translational invariance.  We first examine the periodic system counterparts to establish a reference picture to explore nonlinearity-induced transition. We then determine how this picture changes with increasing quasiperiodicity, with interesting details featuring the interplay of quasiperiodicity, nonlinearity, and topology.  In the strong quasiperiodic regime, pronounced boundary localization can develop well before the topological transition, showing that boundary localization alone cannot be used safely to identify topological edge states. Equally interesting, it is found from this work that the required nonlinearity strength to establish the adiabatic connection between an initial state and topologically protected edge states depends strongly on the initial state.

The remainder of this paper is organized as follows. Section~II introduces a nonlinear SSH model. Section~III presents our real-space numerical method. Section~IV presents our real-space topological analysis of periodic nonlinear systems. Section~V investigates quasiperiodic SSH model, beginning with the phase diagram without nonlinearity and followed by nonlinear results in the weak, intermediate, and strong quasiperiodic regimes. Finally, Sec.~VI summarizes our main findings.

\section{\label{sec:level2}Nonlinear SSH model}
We begin by reviewing the nonlinear Su-Schrieffer-Heeger (SSH) model introduced by Hadad \textit{et al.}~\cite{hadadSelfinducedTopologicalTransitions2016}, one of the earliest demonstrations of a nonlinearity-induced topological phase transition. The dynamics are governed by
\begin{align}
i\partial_{t}\psi_{a,n}
&= J_{1}\psi_{b,n}
+\Bigl[J_{2}+g\left(|\psi_{a,n}|^{2}+|\psi_{b,n-1}|^{2}\right)\Bigr]
\psi_{b,n-1},
\label{eq:1}\\
i\partial_{t}\psi_{b,n}
&= J_{1}\psi_{a,n}
+\Bigl[J_{2}+g\left(|\psi_{b,n}|^{2}+|\psi_{a,n+1}|^{2}\right)\Bigr]
\psi_{a,n+1},
\label{eq:2}
\end{align}
where each unit cell $n$ contains two sublattice sites, denoted by $a$ and $b$, with wavefunction amplitudes $\psi_{a,n}$ and $\psi_{b,n}$. The parameters $J_1$ and $J_2$ represent the intracell and intercell hopping amplitudes of the underlying SSH chain, while $g$ characterizes the strength of a Kerr-type nonlinearity. In the linear limit $g\rightarrow0$, Eqs.~(\ref{eq:1}) and (\ref{eq:2}) reduce to the conventional SSH model~\cite{suSolitonsPolyacetylene1979}.

Assuming Bloch-wave solutions of the form
\begin{align}
\psi_{a,n} &= \Phi_a\, e^{i(kn-Et)},
\label{eq:3}\\
\psi_{b,n} &= \Phi_b\, e^{i(kn-Et)},
\label{eq:4}
\end{align}
where $k$ is the crystal momentum and $E$ is the eigenenergy. Substituting these expressions into Eqs.~(\ref{eq:1}) and (\ref{eq:2}) yields the nonlinear eigenvalue problem
\begin{align}
H(k)
\begin{pmatrix}
\Phi_a\\
\Phi_b
\end{pmatrix}
=
E
\begin{pmatrix}
\Phi_a\\
\Phi_b
\end{pmatrix},
\label{eq:5}
\end{align}
with
\begin{align}
H(k)=
\begin{pmatrix}
0 & J_1+\tilde J_2 e^{-ik}\\
J_1+\tilde J_2 e^{ik} & 0
\end{pmatrix}.
\label{eq:6}
\end{align}
The nonlinear contribution enters through the effective intercell hopping
\begin{align}
\tilde J_2
=
J_2
+
g\left(
|\Phi_a|^2+|\Phi_b|^2
\right),
\label{eq:7}
\end{align}
which depends self-consistently on the wavefunction amplitudes. This amplitude-dependent term renormalizes the intercell coupling and modifies the band structure as the amplitude changes.
The physical origin of the nonlinear topological transition becomes particularly clear when the total amplitude is assumed to be momentum independent and spatially uniform, namely,
\begin{align}
|\Phi_a|^2+|\Phi_b|^2 = I^2.
\label{eq:8}
\end{align} 

Under this assumption, $\tilde{J}_2$ becomes spatially uniform and the nonlinear eigenvalue problem reduces formally to an SSH Hamiltonian with a renormalized intercell hopping. The topological phase boundary is then determined by
\begin{equation}
J_1 = \tilde{J}_2 = J_2 + gI^2,
\label{eq:threshold_boundary}
\end{equation}
which gives the threshold intensity
\begin{equation}
I_{\rm th} = \sqrt{\frac{|J_1 - J_2|}{g}}.
\label{eq:Ith_def}
\end{equation}
For a finite chain, the state is spatially nonuniform and the nonlinear intercell hopping becomes position dependent. Hadad \textit{et al}. studied this finite-chain problem and showed that the nonlinear state evolves continuously from a bulk state to edge-localized states associated with a nonlinearity-induced topological phase.

These results showcase nonlinearity-induced transition from bulk states to topological edge states in a periodic chain, primarily based on momentum-space analysis. To prepare ourselves for quasiperiodic cases treated later,  we first examine here how to directly examine nonlinearity-induced topological phase transitions in real space instead of momentum space. In particular, we follow how nonlinear states and the effective intercell hopping reorganize in real space and how this change is connected to a topological phase transition. We then use our critical insights learned there as a reference to determine how this physical picture is modified by quasiperiodicity.

\section{Method}
\label{sec:method}

To obtain genuine finite-chain eigenstates at the system sizes of interest, we solve the nonlinear eigenvalue problem directly in real space. The finite chain of $N$ dimers is described by the state-dependent equation
\begin{equation}
H(\psi)\,\psi = E\,\psi,
\label{eq:12}
\end{equation}
where the entries of $H$ depend on the state $\psi$ through the nonlinear hopping in Eqs.~\eqref{eq:1}--\eqref{eq:2}. Unlike a linear eigenvalue problem, Eq.~(\ref{eq:12}) cannot be solved by direct diagonalization because the Hamiltonian itself depends on the eigenstate. The task is therefore to find a state that simultaneously satisfies the eigenvalue equation and generates the Hamiltonian appearing in that equation. This self-consistency requirement constitutes the central numerical challenge of the problem.

We solve this problem at fixed norm,
\begin{equation}
\|\psi\|^2 = 1,
\label{eq:fixed_norm_condition}
\end{equation}
and use $g$ as the continuation parameter.  The nonlinear eigenvalue equation and the normalization condition are solved simultaneously as
\begin{equation}
\mathcal{F}(\psi, E) = \begin{pmatrix} H(\psi)\psi - E\psi \\ \|\psi\|^2 - 1 \end{pmatrix} = 0.
\label{eq:newton_system_continuation}
\end{equation}
This formulation fixes the overall amplitude of the state while allowing both the state and its eigenenergy to evolve with $g$.

The continuation begins at $g=0$, where we choose the smallest-positive-energy eigenstate of the corresponding linear chain. The nonlinear strength is then increased incrementally. At each value of $g$, the converged solution from the previous step is used as the initial state for the next calculation. In this way, the same nonlinear branch is followed continuously as $g$ is increased. 
At each continuation step, the nonlinear eigenvalue equation and the normalization condition are solved simultaneously using Newton iteration with the analytical Jacobian~\cite{kelley2003solving}. The state and eigenenergy are updated together until the numerical residual reaches a tolerance of $10^{-10}$. 


\section{Periodic Case: Validation of the Real-Space Method}
\label{sec:periodic}

\subsection{Real-space changes of nonlinear eigenstates}

To understand a nonlinear topological transition in real space, we first examine how the nonlinear eigenstates evolve in the case of  periodic chains (absence of quasiperiodicity).  This provides a real-space picture of the transition discussed in Sec. II.

For a stationary state, we write
\begin{equation}
\psi_{a,n}(t)=a_n e^{-iEt}, \qquad \psi_{b,n}(t)=b_n e^{-iEt}.
\end{equation}
The effective intracell and intercell hoppings are then
\begin{equation}
t_{1,n}=J_1,
\end{equation}
and
\begin{equation}
t_{2,n}=J_2+g\rho_n,
\label{Eq:t_2n}
\end{equation}
where
\begin{equation}
\rho_n=|b_n|^2+|a_{n+1}|^2
\end{equation}
is the probability on the two sites connected by the $n$-th intercell bond. Although the bare hoppings $J_1$ and $J_2$ are spatially uniform, the nonlinear state therefore generates a position-dependent intercell hopping through $\rho_n$. In the following analysis, we assume the underlying linear system is topologically trivial, namely $J_1 > J_2$.

To characterize the spatial distribution of the state, we also define the probability in unit cell $n$,
\begin{equation}
P_n=|a_n|^2+|b_n|^2,
\end{equation}
with $\sum_n P_n=1$. The intensity is then
\begin{equation}
I^2=\max_n P_n.
\label{Eq:Intensity}
\end{equation}

At $g=0$, the intercell hopping is uniform, $t_{2,n}=J_2$, and the system is in the smallest-positive-energy state of the finite open SSH chain. The bond density $\rho_n$ of this state has a single broad maximum around the center of the chain. To determine where the nonlinear modification first develops as $g$ is increased from zero, we differentiate Eq.~\eqref{Eq:t_2n}, which gives
\begin{equation}
\left.\frac{\partial t_{2,n}}{\partial g}\right|_{g=0} = \rho_n(0).
\label{eq:dt2_dg_g0}
\end{equation}
Thus, the initial change of the intercell hopping is largest around the center, where the bond density of the linear state is the largest.

\begin{figure*}[t]
\centering
\includegraphics[width=\textwidth]{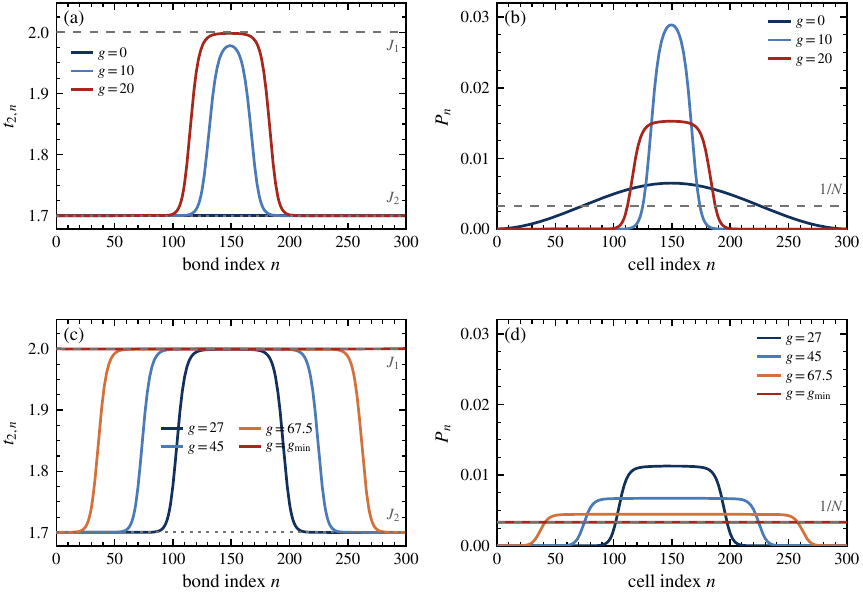}
\caption{Periodic case: Real-space profiles of nonlinear eigenstates for $N=300$, $J_1=2$, and $J_2=1.7$. (a,b)~Change of the effective intercell hopping $t_{2,n}=J_2+g\rho_n$ and cell probability $P_n$ for relatively small $g$. The broad linear state first becomes concentrated around the center and subsequently begins to re-expand. (c,d)~Features of nonlinear states showing the outward expansion in the region near-$J_1$  and the corresponding widening and lowering of $P_n$. Near $g_{\min}$, the region extends across essentially the whole chain and $P_n$ approaches a nearly uniform profile. In panels~(a) and~(c), the dashed and dotted lines indicate $J_1$ and $J_2$, respectively. In panels~(b) and~(d), the dashed line indicates $1/N$.}
\label{fig:Fig1_Periodic}
\end{figure*}

Figure~\ref{fig:Fig1_Periodic} shows how the effective intercell hoppings $t_{2,n}$ and the adiabatically connected eigenstates change as the nonlinear strength is increased. Starting from the broad linear state at $g=0$, the state first becomes more concentrated around the center of the chain [Figs.~\ref{fig:Fig1_Periodic}(a) and \ref{fig:Fig1_Periodic}(b)]. With increasing $g$, this trend reverses and the state begins to broaden. A region where $t_{2,n}$ lies close to $J_1$ then develops around the center and expands toward the boundaries, while the exterior remains close to $J_2$ [Fig.~\ref{fig:Fig1_Periodic}(c)]. The corresponding probability profile becomes wider and lower as this region expands [Fig.~\ref{fig:Fig1_Periodic}(d)]. We define $g = g_{\min}$ to be the nonlinear parameter when the near-$J_1$ region extends across essentially the whole chain and $P_n$ approaches a nearly uniform profile. Thus, the nonlinear eigenstates change from the broad linear state through an initial central concentration, followed by re-expansion across the chain as $g$ is increased.

The significance of the near-$J_1$ region can be understood from the zero-energy real-space recurrence. As this region develops and expands, the eigenenergy becomes small compared with $J_1$, as shown in Appendix~\ref{app:energy}. At zero energy,

\begin{equation}
a_{n+1} = -\frac{J_1}{t_{2,n}} a_n.
\label{eq:zero_energy_recurrence_periodic}
\end{equation}
For $t_{2,n} < J_1$, $|a_{n+1}| > |a_n|$, whereas for $t_{2,n} > J_1$, $|a_{n+1}| < |a_n|$. The condition $t_{2,n} = J_1$ therefore separates the locally growing and decaying behaviors. The expanding near-$J_1$ region seen in Fig.~\ref{fig:Fig1_Periodic} approaches this equal-hopping condition as $g$ is increased, while the corresponding probability profile broadens across the same region.

Outside the near-$J_1$ region, the probability and bond density become small, and the effective intercell hopping approaches its bare value $J_2$. The hopping profile therefore consists of an expanding near-$J_1$ region surrounded by $J_2$-like exterior regions.

\subsection{Expansion of the near-$J_1$ region and topological transition}

\begin{figure}[t]
\centering
\includegraphics[width=\linewidth]{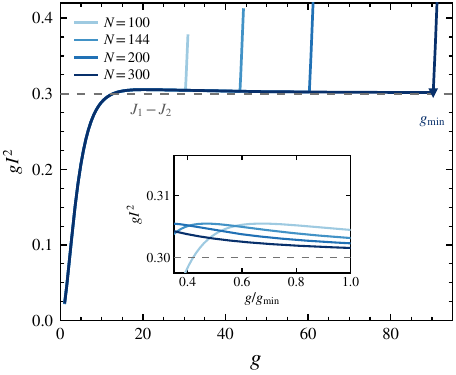}
\caption{Periodic case: $gI^2$ as a function of $g$ for $J_1=2$, $J_2=1.7$, and different system sizes $N$. The dashed line marks $J_1-J_2=0.3$. With increasing $N$, the plateau approaches this value. The inset shows the same quantity as a function of $g/g_{\min}$.}
\label{fig:Fig2_Periodic}
\end{figure}

We next examine how the expansion of the near-$J_1$ region identified in Fig.~\ref{fig:Fig1_Periodic} is related to the redistribution of probability across the chain. Within this region, $t_{2,n} \simeq J_1$, and Eq.~\eqref{Eq:t_2n} gives
\begin{equation}
\rho_n \simeq \frac{J_1 - J_2}{g}.
\label{eq:rhon_plateau_approx}
\end{equation}
Thus, the bond density in this region decreases approximately as $1/g$, while the nonlinear contribution $g\rho_n$ remains close to $J_1 - J_2$.

We next relate the bond density to the cell probability. From their definitions,
\begin{equation}
\rho_n - P_n = |a_{n+1}|^2 - |a_n|^2.
\label{eq:rho_minus_P_diff}
\end{equation}
Within the near-$J_1$ region, Eq.~\eqref{eq:zero_energy_recurrence_periodic} gives $|a_{n+1}| \simeq |a_n|$. The difference between $\rho_n$ and $P_n$ therefore becomes small, and hence
\begin{equation}
P_n \simeq \rho_n \simeq \frac{J_1 - J_2}{g}.
\label{eq:Pn_rhon_approx_plateau}
\end{equation}

This relation explains the changes in the probability profile observed in Fig.~\ref{fig:Fig1_Periodic}(d). As $g$ increases, $P_n$ within the near-$J_1$ region decreases approximately as $1/g$. Since the total probability remains normalized, the profile must extend over an increasing number of cells. Let $W$ denote the width of the nearly flat probability profile. Normalization gives
\begin{equation}
1 \simeq W P_n,
\label{eq:norm_W_Pn_periodic}
\end{equation}
and therefore
\begin{equation}
W \simeq \frac{g}{J_1 - J_2}.
\label{eq:W_scaling_periodic}
\end{equation}
Thus, the widening and lowering of the probability profile seen in Fig.~\ref{fig:Fig1_Periodic}(d) are two consequences of the same real space changes of the nonlinear eigenstates. Its width grows approximately as $g$, whereas its height decreases approximately as $1/g$.

Since $I^2$ is the maximum cell probability, the nearly flat profile gives
\begin{equation}
I^2 \simeq \frac{J_1 - J_2}{g},
\label{eq:I2_plateau_scaling}
\end{equation}
and consequently
\begin{equation}
gI^2 \simeq J_1 - J_2.
\label{eq:gI2_plateau_constant}
\end{equation}
This behavior can persist only while the near-$J_1$ region continues to expand. In a finite chain, the expansion is limited by the system size and eventually extends over all $N$ cells. Since the total probability is fixed, the maximum cell probability satisfies
\begin{equation}
I^2 \ge \frac{1}{N},
\label{eq:I2_uniform_bound}
\end{equation}
with equality only for a uniform probability distribution. As the near-$J_1$ region approaches the size of the chain, the intensity therefore approaches its minimum value,
\begin{equation}
I_{\min}^2 \to \frac{1}{N}.
\label{eq:Imin2_asymptotic_limit}
\end{equation}
At the same time, $W \to N$. Using the relation obtained above for $W$, we obtain
\begin{equation}
\frac{g_{\min}}{N} \to J_1 - J_2.
\label{eq:gmin_over_N_limit}
\end{equation}

\begin{figure*}[t]
\centering
\includegraphics[width=\textwidth]{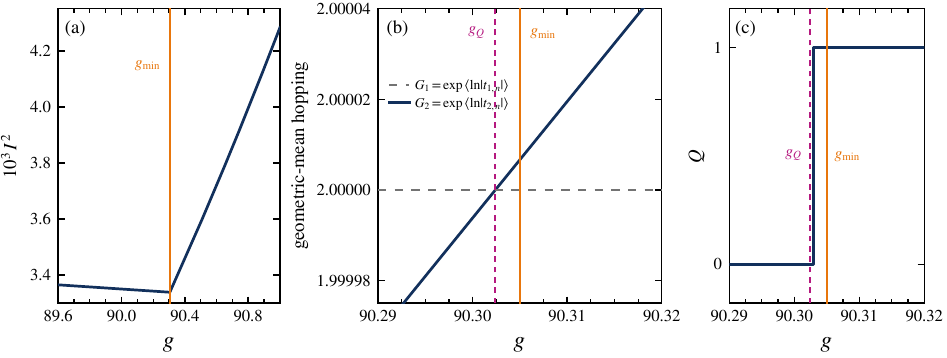}
\caption{Periodic case: Intensity minimum and topological transition with $J_1=2$, $J_2=1.7$, and $N=300$. (a)~Intensity $I^2$ near its minimum at $g_{\min}$. (b)~Geometric-mean hopping scales $G_1$ and $G_2$, which become equal at $g_Q$. (c)~Real-space topological invariant $Q$, which changes from 0 to 1 at $g_Q$. The magenta dashed and orange solid lines mark $g_Q$ and $g_{\min}$, respectively.}
\label{fig:Fig3_Periodic}
\end{figure*}

Figures~\ref{fig:Fig2_Periodic} and \ref{fig:Fig3_Periodic}(a) show the corresponding numerical behavior. Figure~\ref{fig:Fig2_Periodic} shows that $gI^2$ develops a broad plateau near $J_1 - J_2 = 0.3$. As the system size increases, the plateau approaches this value more closely and extends to larger $g$. Its endpoint also moves approximately in proportion to $N$, consistent with the relation $g_{\min}/N \to J_1 - J_2$ obtained above. The inset shows the same result as a function of $g/g_{\min}$, where the curves approach the plateau value as the endpoint is reached. Figure~\ref{fig:Fig3_Periodic}(a) resolves the intensity near this endpoint for $N=300$. The intensity decreases up to $g_{\min}$ and increases beyond it, showing that the endpoint of the expanding near-$J_1$ region coincides with the minimum of $I^2$. Together with the nearly uniform probability profile near $g_{\min}$ in Fig.~\ref{fig:Fig1_Periodic}(d), these results support the real space picture developed above.

Having identified $g_{\min}$ with the point where the near-$J_1$ region extends across essentially the whole chain, we next examine its relation to the topological transition. We characterize the topology using the reflection matrix invariant \cite{Fulga_Hassler_Akhmerov_Beenakker_2011},
\begin{equation}
Q = \frac{1}{2} \left[ 1 - \operatorname{sgn}\left( \prod_{n=1}^{N-1} (-t_{1,n})^2 - \prod_{n=1}^{N-1} (-t_{2,n})^2 \right) \right],
\label{eq:reflection_matrix_invariant}
\end{equation}
where $Q=0$ and $Q=1$ correspond to the trivial and topological phases, respectively. The transition occurs when
\begin{equation}
\prod_{n=1}^{N-1} |t_{1,n}|^2 = \prod_{n=1}^{N-1} |t_{2,n}|^2.
\label{eq:prod_t1_prod_t2}
\end{equation}
where the products run over the $N-1$ bonds because of the open boundary condition.
Taking the logarithm gives
\begin{equation}
\langle \ln |t_{1,n}| \rangle = \langle \ln |t_{2,n}| \rangle.
\label{eq:log_hopping_mean_equality}
\end{equation}
We therefore define the geometric-mean hopping scales
\begin{equation}
G_1 = \exp\langle \ln |t_{1,n}| \rangle, \quad G_2 = \exp\langle \ln |t_{2,n}| \rangle,
\label{eq:G1_G2_defs}
\end{equation}
so that the topological transition is determined by
\begin{equation}
G_2 = G_1.
\label{eq:G2_G1_topological_condition}
\end{equation}
Equation~\eqref{eq:reflection_matrix_invariant} is the invariant we use to identify the topological phase throughout this work. The equivalent condition $G_2 =G_1$ is simply a more convenient form of the same criterion, and we quote whichever is more transparent in a given figure.
For the periodic chain, $t_{1,n} = J_1$, and therefore $G_1 = J_1$. The connection with the real space changes can be seen from Fig.~\ref{fig:Fig1_Periodic}(c). As the near-$J_1$ region expands across the chain, $t_{2,n}$ approaches $J_1$ over an increasing number of bonds. Consequently, $G_2$ approaches $G_1$.

Figure~\ref{fig:Fig3_Periodic}(b) shows the two geometric mean hopping scales near $g_{\min}$. At nonlinear parameter $g_Q$, $G_2 = G_1$ and the real space topological invariant changes from $Q = 0$ to $Q = 1$, as shown in Fig.~\ref{fig:Fig3_Periodic}(c). For $N = 300$, we find
\begin{equation}
    g_Q \simeq 90.3024, \quad g_{\min} \simeq 90.3051.
\end{equation}
Thus, the topological transition occurs very close to, but not exactly at, the intensity minimum for a finite chain. The difference $g_Q - g_{\min}$ approaches zero with increasing $N$, as shown in Appendix~\ref{sec:finite-size-gQ-gmin}.

The two nonlinear scales therefore have distinct real space origins. The intensity minimum $g_{\min}$ occurs when the expanding near-$J_1$ region reaches essentially the whole chain, while the topological transition $g_Q$ is determined by the global condition $G_2 = G_1$. Their difference vanishes with increasing system size, and the two conditions become asymptotically coincident.

\begin{figure}[t]
\centering
\includegraphics[width=\linewidth]{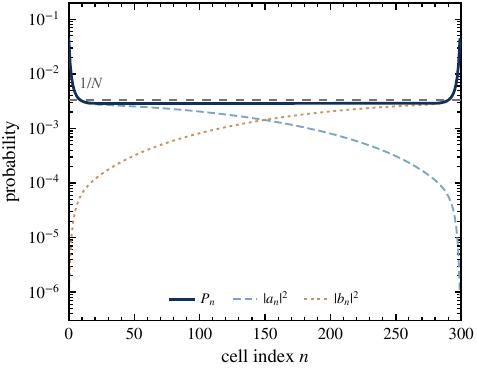}
\caption{Periodic case: Nonlinear edge state for $N = 300$, $J_1 = 2$, $J_2 = 1.7$, and $g = 105$, beyond both $g_{\min}$ and $g_Q$. The cell probability $P_n$ and sublattice probabilities $|a_n|^2$ and $|b_n|^2$ show the two-edge profile, with the two sublattices predominantly localized at opposite boundaries.}
\label{fig:Fig4_Periodic}
\end{figure}

\subsection{Development of  nonlinear edge states}

Beyond the transition at $g_Q$, the geometric mean hopping scales satisfy
\begin{equation}
    G_2 > G_1,
\end{equation}
and the real space topological invariant is $Q = 1$. We now examine how the nonlinear state evolves in this topological regime.

Figure~\ref{fig:Fig4_Periodic} shows the probability profile at $g = 105$, beyond both $g_{\min}$ and the topological transition at $g_Q$. In contrast to the nearly uniform profile near $g_{\min}$, the probability is now enhanced near both boundaries and decreases toward the interior. The two sublattices are predominantly localized at opposite ends of the chain, giving the two edge profile characteristic of the nonlinear state beyond the transition.

The results so far for the periodic chain as our reference point can therefore be summarized as follows. The near-$J_1$ region first expands across the chain and the intensity reaches its minimum at $g_{\min}$. The topological transition occurs at $g_Q$, where $G_2 = G_1$ and $Q$ changes from $0$ to $1$. The two nonlinear scales become asymptotically coincident with increasing system size. With further increase of $g$, probability accumulates near the boundaries and the state evolves into the two edge profile shown in Fig.~\ref{fig:Fig4_Periodic}.

\section{Introduction of quasiperiodic potential}\label{sec:level3}

\begin{figure}[t]
\centering
\includegraphics[width=\linewidth]{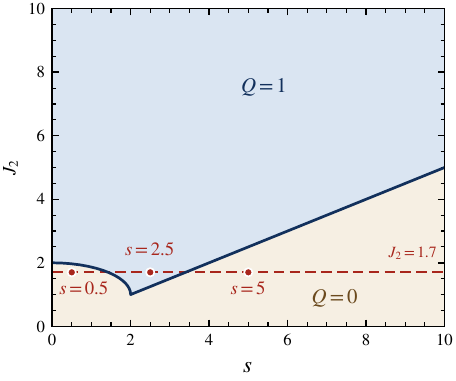}
\caption{Linear quasiperiodic case: Phase diagram for $J_1 = 2$. The solid curve shows the analytical phase boundary $J_2^c(s) = G_1$ [Eq.~\eqref{eq:J2c_is_G1}], separating the topological $Q = 1$ and trivial $Q = 0$ phases. The horizontal dashed line marks $J_2 = 1.7$, with $s = 0.5$, $2.5$, and $5$ indicating the weak, intermediate, and strong quasiperiodic regimes considered below.}
\label{fig:Fig5_Periodic}
\end{figure}

Having better understood the real space changes of nonlinear eigenstates as nonlinearity strength increases and their connections to topological phase transitions in periodic chains, we now turn to  quasiperiodic systems. We first examine how quasiperiodicity modifies the topology of the underlying linear system. This provides the reference for our investigation of the nonlinear eigenstates below.

We consider an SSH chain consisting of $N$ dimers, with a quasiperiodic modulation applied to the intracell hopping. The dynamics are governed by
\begin{align}
i\frac{\partial\psi_{a,n}}{\partial t} &= \left[J_1+s\cos(2\pi\beta n)\right]\psi_{b,n} + J_2\psi_{b,n-1}, \label{eq:linear_qp_dyn_a} \\
i\frac{\partial\psi_{b,n}}{\partial t} &= \left[J_1+s\cos(2\pi\beta n)\right]\psi_{a,n} + J_2\psi_{a,n+1}, \label{eq:linear_qp_dyn_b}
\end{align}
where $s$ characterizes the strength of the quasiperiodic modulation and $\beta=(\sqrt{5}+1)/2$. The quasiperiodic modulation acts on the intracell hopping, while the intercell hopping remains uniform. In the limit $s\to 0$, the model reduces to the conventional SSH chain.

Using the notation introduced in Sec.~IV, the effective hoppings in the linear quasiperiodic system are
\begin{equation}
t_{1,n} = J_1 + s\cos(2\pi\beta n),
\label{eq:t1_lin_qp}
\end{equation}
and
\begin{equation}
t_{2,n} = J_2.
\label{eq:t2_lin_qp}
\end{equation}
Because the quasiperiodic modulation breaks translational invariance, the conventional momentum-space winding number is unavailable. Being a real-space quantity, $Q$ is unaffected and is what identifies the phases in Fig.~\ref{fig:Fig5_Periodic}. For the linear system, the phase boundary can in addition be written in closed form, obtained from the localization length of the zero-energy mode. Following Ref.~\cite{miranda_mechanical_2024}, the corresponding Lyapunov exponent is
\begin{equation}
\gamma =
\begin{cases}
\displaystyle \ln\left(\frac{J_1+\sqrt{J_1^2-s^2}}{2J_2}\right), & J_1 > s, \\[3mm]
\displaystyle \ln\left(\frac{s}{2J_2}\right), & J_1 < s.
\end{cases}
\label{eq:lyapunov_exponent}
\end{equation}
The transition occurs at $\gamma=0$, which gives
\begin{equation}
J_2^c(s) =
\begin{cases}
\displaystyle \frac{J_1+\sqrt{J_1^2-s^2}}{2}, & J_1 > s, \\[3mm]
\displaystyle \frac{s}{2}, & J_1 < s.
\end{cases}
\label{eq:J2c_analytical}
\end{equation}
This critical value is the geometric mean of the quasiperiodic intracell hopping. Using the geometric-mean hopping scales introduced in Sec.~IV, we have
\begin{equation}
J_2^c(s) = G_1.
\label{eq:J2c_is_G1}
\end{equation}
Since $t_{2,n}=J_2$ is uniform in the linear system,
\begin{equation}
G_2 = J_2.
\label{eq:G2_is_J2_lin}
\end{equation}
The linear topological transition is therefore determined by
\begin{equation}
G_2 = G_1.
\label{eq:lin_trans_G2_G1}
\end{equation}

For $J_1 = 2$ this boundary is the solid curve in Fig.~\ref{fig:Fig5_Periodic}. Throughout the nonlinear analysis, we fix $J_2 = 1.7$. Along this cut, increasing $s$ takes the linear system from a trivial phase into a topological phase and then back into a trivial phase. We choose $s = 0.5$, $2.5$, and $5$ as representative points in these three regions. They define the weak, intermediate, and strong quasiperiodic regimes considered below.

\subsection{Weak quasiperiodic regime}

\begin{figure}[t]
\centering
\includegraphics[width=\linewidth]{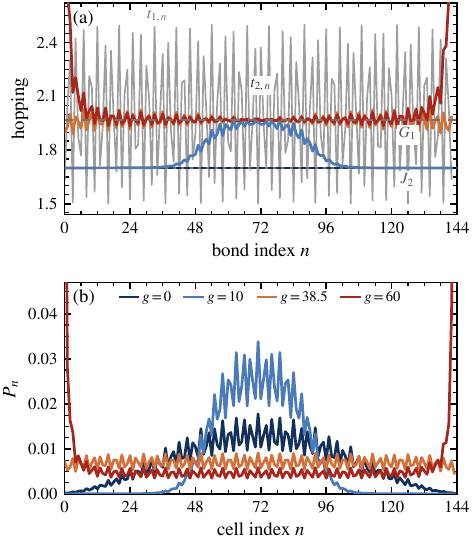}
\caption{Weak quasiperiodic case: Real-space changes for $s = 0.5$, $J_1 = 2$, $J_2 = 1.7$, and $N = 144$. (a)~Quasiperiodic intracell hopping $t_{1,n}$ and nonlinear intercell hopping $t_{2,n}$ for increasing $g$. The horizontal dashed and dotted lines indicate $G_1$ and $J_2$, respectively. (b)~Corresponding cell probability $P_n$. The colors from dark blue to red indicate increasing $g$.}
\label{fig:Fig6_Weak_QP}
\end{figure}

\begin{figure*}[t]
\centering
\includegraphics[width=\linewidth]{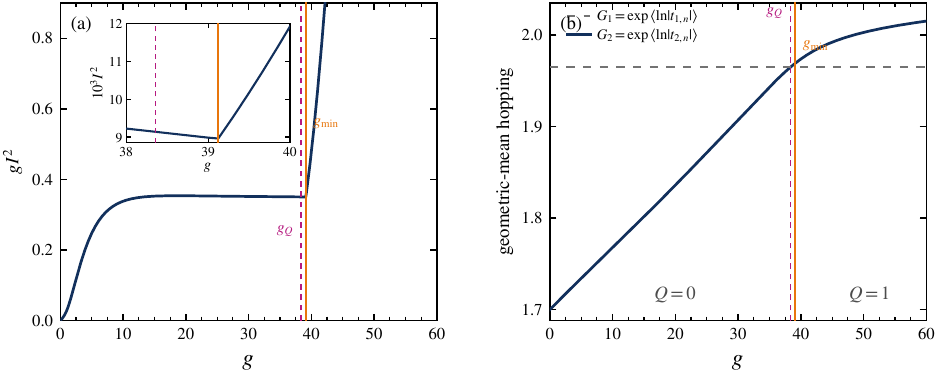}
\caption{Weak quasiperiodic case: Intensity and topological transition for $s = 0.5$, $J_1 = 2$, $J_2 = 1.7$, and $N = 144$. (a)~$g I^2$ as a function of $g$. The inset shows the intensity $I^2$ near its minimum at $g_{\min}$. (b)~Geometric mean hopping scales $G_1$ and $G_2$, which become equal at $g_Q$. The regions $Q = 0$ and $Q = 1$ are indicated. The magenta dashed and orange solid lines mark $g_Q$ and $g_{\min}$, respectively.}
\label{fig:Fig7_Weak_QP}
\end{figure*}

We first consider the weak quasiperiodic regime, represented by $s=0.5$. For $J_1=2$ and $J_2=1.7$, the linear system is topologically trivial. The intracell hopping is quasiperiodic,
\begin{equation}
t_{1,n} = J_1 + s\cos(2\pi\beta n),
\label{eq:t1_weak_sec}
\end{equation}
while the nonlinear intercell hopping is
\begin{equation}
t_{2,n} = J_2 + g\rho_n.
\label{eq:t2_weak_sec}
\end{equation}

Figure~\ref{fig:Fig6_Weak_QP}(a) shows how the intercell hopping evolves as $g$ is increased. At $g = 0$, $t_{2,n} = J_2$ is uniform, while $t_{1,n}$ retains the quasiperiodic modulation. The nonlinear correction first develops predominantly around the center of the chain. With increasing $g$, $t_{2,n}$ rises over an increasing part of the chain and extends across almost the whole system near $g = 38.5$. In contrast to the periodic case, $t_{2,n}$ does not approach $t_{1,n}$ bond by bond. Instead, the nonlinear intercell hopping remains much smoother than the quasiperiodic intracell hopping and approaches the global scale $G_1$. At larger $g$, $t_{2,n}$ becomes strongly enhanced near the boundaries.

The corresponding probability profiles are shown in Fig.~\ref{fig:Fig6_Weak_QP}(b). The state first becomes more concentrated around the center, where the nonlinear modification of $t_{2,n}$ initially develops. As $t_{2,n}$ extends across the chain, the probability profile also broadens. At larger $g$, the probability becomes strongly enhanced near the boundaries together with the boundary enhancement of $t_{2,n}$.

Despite this spatial modulation, $g I^2$ develops a broad plateau, as shown in Fig.~\ref{fig:Fig7_Weak_QP}(a). However, its origin differs from the periodic case. The probability profile remains nonuniform, and the relation $P_n \simeq \rho_n$ used to obtain the periodic plateau no longer holds. The value of the plateau therefore cannot be obtained from the periodic relation $g I^2 \simeq J_1 - J_2$ and we do not have a corresponding prediction for the plateau in the quasiperiodic case.

The inset of Fig.~\ref{fig:Fig7_Weak_QP}(a) shows that the intensity reaches its minimum at $g_{\min} \simeq 39.122$.
Unlike the periodic case, the probability profile remains nonuniform near $g_{\min}$. The minimum instead results from a change in the cell that determines $I^2$. Below $g_{\min}$, $I^2$ is determined by a bulk cell. As $g$ increases, the probability at this cell decreases while the probability at the left boundary increases. Near $g_{\min}$, the two probabilities cross and the cell determining $I^2$ changes from the bulk to the left boundary, as shown in Appendix~\ref{app:weakmin}. After $g_{\min}$, as $g$ increases, the probabilities of both boundaries then increase monotonically. Hence, $g_{\min}$ estimates the nonlinear parameter when boundary localization starts to occur.

Next, we determine the topological transition from the global hopping scales introduced in Eq.~\eqref{eq:G1_G2_defs}. Figure~\ref{fig:Fig7_Weak_QP}(b) shows $G_1$ and $G_2$ as $g$ is increased. The intracell scale $G_1$ remains fixed because the quasiperiodic modulation is independent of $g$, while $G_2$ increases as the nonlinear contribution modifies $t_{2,n}$. The two scales become equal at $ g_Q \simeq 38.3593.$
At the same point, the real space invariant changes from $Q = 0$ to $Q = 1$. The weak quasiperiodic system therefore undergoes a nonlinearity-induced topological transition determined by $G_2 = G_1$.

Thus, in the weak quasiperiodic regime, increasing the nonlinear strength drives the initially trivial system into a topological phase. The transition occurs at $g_Q$, where $G_2 = G_1$ marks the change from $Q = 0$ to $Q = 1$. The intensity reaches its minimum at the nearby but distinct $g_{\min}$ for the finite system considered here. Unlike the periodic case, this minimum results from a change in the cell that determines $I^2$, from a bulk cell to the left boundary, rather than from the probability profile becoming nearly uniform across the chain. We note that in this regime, $g_{\min}$ is consistently close to $g_Q$, which indicates that boundary localization corresponds to a topological transition.


\begin{figure}[b]
\centering
\includegraphics[width=\linewidth]{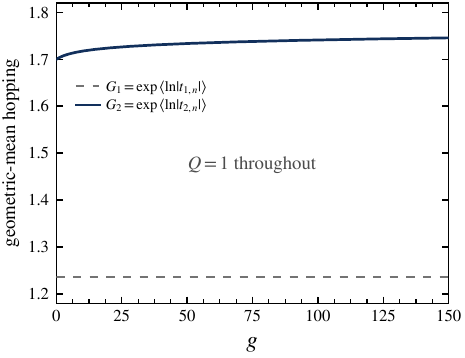}
\caption{Intermediate quasiperiodic case: Geometric mean hopping scales for $s = 2.5$, $J_1 = 2$, $J_2 = 1.7$, and $N = 144$. The solid line shows $G_2$, while the dashed line shows $G_1$. The geometric mean of the nonlinear intercell hopping remains above that of the quasiperiodic intracell hopping throughout the range considered, corresponding to $Q = 1$.}
\label{fig:Fig8_Intermediate_QP}
\end{figure}

\subsection{Intermediate quasiperiodic regime}

We next consider the intermediate quasiperiodic regime with $s=2.5$. In this case, the linear system is already topological. For the finite $N=144$ chain, the geometric mean of the intracell hopping is
\begin{equation}
G_1 \simeq 1.236.
\label{eq:G1_intermediate_val}
\end{equation}
At $g=0$, the intercell hopping is uniform and therefore
\begin{equation}
G_2(0) = J_2 = 1.7.
\label{eq:G2_intermediate_initial}
\end{equation}
Thus, the condition
\begin{equation}
G_2(0) < G_1,
\label{eq:G2_gt_G1_interm}
\end{equation}
is never reached, and the real space invariant remains $Q=1$. We therefore examine whether increasing the nonlinear strength can drive the system out of this initially topological phase.

Figure~\ref{fig:Fig8_Intermediate_QP} shows the changes in the geometric mean hopping scales as $g$ is increased. The intracell scale $G_1$ remains fixed, while $G_2$ remains above $G_1$ throughout the range considered. The condition $G_2 = G_1$ is therefore never reached, and the real space invariant remains $Q = 1$.
\begin{figure*}[t]
\centering
\includegraphics[width=\textwidth]{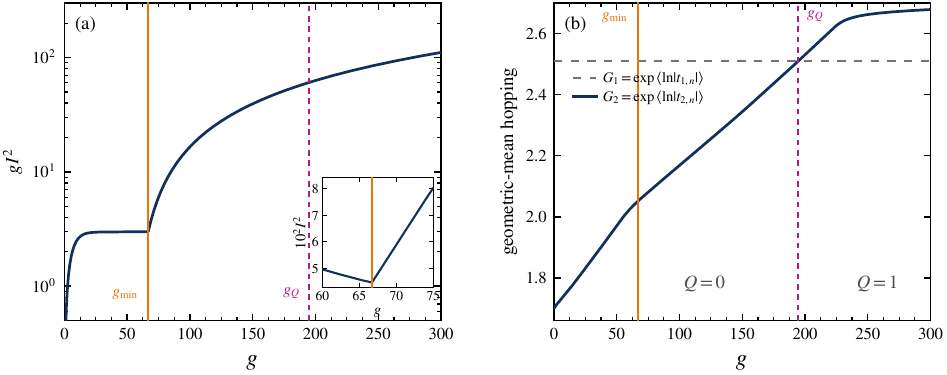}
\caption{Strong quasiperiodic case: Intensity and topological transition for $s = 5$, $J_1 = 2$, $J_2 = 1.7$, and $N = 144$. (a)~$g I^2$ as a function of $g$. The inset shows the intensity $I^2$ near its minimum at $g_{\min}$. (b)~Geometric mean hopping scales $G_1$ and $G_2$, which become equal at $g_Q$. The regions $Q = 0$ and $Q = 1$ are indicated. The magenta dashed and orange solid lines mark $g_Q$ and $g_{\min}$, respectively.}
\label{fig:Fig9_Strong_QP}
\end{figure*}

The intermediate regime therefore provides a distinct regime. The system is already topological in the linear limit and remains in the same topological phase as the nonlinear strength is increased. No nonlinearity induced topological transition occurs over the range considered.

\subsection{Strong quasiperiodic regime}

We finally consider the strong quasiperiodic regime with $s = 5$. The corresponding linear system is topologically trivial. We choose the lowest positive energy state at $g=0$, which is strongly localized around a bulk cell by the quasiperiodic hopping profile. We examine how this initially localized state evolves as the nonlinear strength is increased.

Figure~\ref{fig:Fig9_Strong_QP}(a) shows the corresponding changes of $gI^2$. The intensity reaches its minimum at
$g_{\min} \simeq 66.682.$
As in the weak quasiperiodic case, this minimum results from a change in the cell that determines $I^2$. Below $g_{\min}$, $I^2$ is determined by the quasiperiodically localized bulk cell. As $g$ increases, the probability at this cell decreases while the probability at the left boundary increases. Near $g_{\min}$, the two probabilities cross and the cell determining $I^2$ changes from the bulk to the left boundary.

We next examine how the topology evolves as the nonlinear strength is increased. At $g_{\min}$, Fig.~\ref{fig:Fig9_Strong_QP}(b) shows that $G_2$ remains well below $G_1$, while the real space invariant remains $Q = 0$. The intensity minimum therefore does not correspond to a topological transition. With further increase of $g$, $G_2$ continues to increase and reaches $G_1$ at $g_Q \simeq 194.737$. At the same point, $Q$ changes from $0$ to $1$. Thus, nonlinearity does induce a topological transition in the strong quasiperiodic regime, but at a substantially larger nonlinear strength than $g_{\min}$.
\begin{figure*}[t]
\centering
\includegraphics[width=\textwidth]{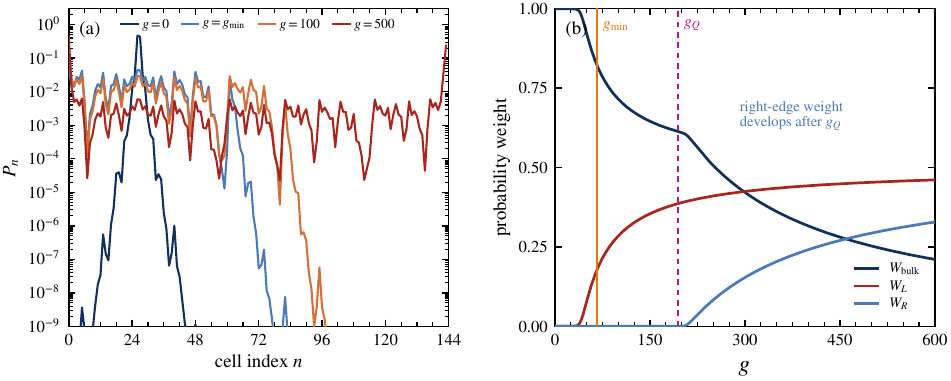}
\caption{Strong quasiperiodic case: Real-space changes for $s = 5$, $J_1 = 2$, $J_2 = 1.7$, and $N = 144$. (a) Cell probability $P_n$ for representative values of $g$. The colors from dark blue to red indicate increasing $g$. (b) Probability weights in the bulk $W_{\mathrm{bulk}}$ and at the two boundaries $W_L$ and $W_R$ as a function of $g$. The boundary weights are defined over the first and last ten cells, with the remaining cells defining the bulk. The magenta dashed and orange solid lines mark $g_Q$ and $g_{\min}$, respectively.}
\label{fig:Fig10_Strong_QP}
\end{figure*}

To understand how the state evolves across this large separation between $g_{\min}$ and $g_Q$, Fig.~\ref{fig:Fig10_Strong_QP} shows the redistribution of probability as $g$ is increased across $g_{\min}$ and $g_Q$.
The real space changes across these two nonlinear strengths are shown in Fig.~\ref{fig:Fig10_Strong_QP}. Figure~\ref{fig:Fig10_Strong_QP}(b) shows that the probability is initially concentrated in the bulk. As $g$ increases, the bulk weight decreases while the left boundary weight increases continuously. As in the weak quasiperiodic regime, probability develops near the left boundary before $g_{\min}$, and the intensity minimum occurs when the cell determining $I^2$ changes from the quasiperiodically localized bulk cell to the left boundary. The left boundary remains dominant as $g$ is increased toward $g_Q$, while the right boundary weight remains small. Beyond $g_Q$, the right boundary weight begins to increase and continues to develop with increasing $g$. At larger $g$, probability accumulates near both boundaries, giving the two-edge profile shown at $g = 500$ in Fig.~\ref{fig:Fig10_Strong_QP}(a).

Thus, at $s=5$ substantial boundary localization develops before the initially trivial system enters the topological phase, and boundary localization therefore does not by itself identify the topological transition. Both nonlinear scales $g_{\min}$ and $g_Q$ measure how far the state must spread, but to different boundaries. The intensity minimum occurs when the probability at the nearer boundary overtakes that of the localized bulk cell, so $g_{\min}$ is set by the distance $d$ from the localization center to that boundary. The topological transition instead requires $G_2 = G_1$, a geometric mean over all bonds, so the state must also reach the farther boundary, and $g_Q$ is set by that distance $N-d-1$. Both dependencies are verified independently in Appendix~\ref{app:separation}, which also shows the same behavior for other initial states and for other values of $s$ in the strong quasiperiodic regime. The separation is therefore dependent on the difference of the two distances. It is large when the initial state lies close to one boundary and small when it lies near the center, where the two distances coincide.


At larger nonlinear strength, probability develops at both boundaries and the two-edge profile emerges within the topological phase.

\section{Conclusion}

In this work, we have investigated how nonlinearity drives topological transitions directly in real space. In a periodic chain, our real space treatment allows us to see clear transitions in nonlinear eigenstates, with the eigenstate and the effective hopping reorganizing progressively across the chain. Quasiperiodicity then changes this picture in a nontrivial way. In the weak quasiperiodic regime, nonlinearity still drives an initially trivial system into a topological phase, although the intensity minimum required no longer has the same origin as in the periodic system. In the regime of an intermediate quasiperiodic potential, if the linear system is already topologically nontrivial,  it is found to remain so as the nonlinear strength is increased. In the strong quasiperiodic regime, nonlinearity again drives an initially trivial system into a topological phase, but the transition can occur at a nonlinear strength that is substantially larger than that expected from an intensity analysis only suitable for weak and intermediate quasiperiodicity regimes. This shows that boundary localization alone is not a safe signature of topological edge states in nonlinear quasiperiodic systems. As another important message from this work,  the required nonlinearity strength to establish the adiabatic connection between an initial state and topologically protected edge states depends strongly on the initial state. 
The real space treatment developed and advocated here therefore provides a useful basis for investigating nonlinear topological transitions beyond translationally invariant systems and for the preparation of nonlinear topological edge states.  An important next question is how this picture extends to systems without chiral symmetry.

\begin{acknowledgments}
  J.G. acknowledges support by the National Research Foundation, Singapore, through the National Quantum Office, hosted in A*STAR, under its Centre for Quantum Technologies Funding Initiative (S24Q2d0009).  
\end{acknowledgments}

\appendix

\begin{figure}[t]
\centering
\includegraphics[width=\linewidth]{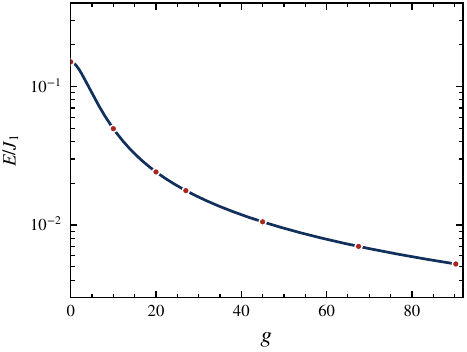}
\caption{Periodic case: Eigenenergy $E/J_1$ as a function of $g$ for $N = 300$, $J_1 = 2$, and $J_2 = 1.7$. The marked points correspond to the values of $g$ shown in Fig.~\ref{fig:Fig1_Periodic}. The eigenenergy decreases as the near-$J_1$ region expands and becomes small near $g_{\min}$.}
\label{fig:App_Fig1}
\end{figure}

\section{Eigenenergy in periodic systems}
\label{app:energy}
In the periodic system, the eigenenergy decreases as the nonlinear strength is increased. Figure~\ref{fig:App_Fig1} shows $E/J_1$ for $N=300$, using the same parameters as in Fig.~\ref{fig:Fig1_Periodic}. As the near-$J_1$ region develops and expands across the chain, $E/J_1$ decreases continuously and reaches $5.2 \times 10^{-3}$ near $g_{\min}$. The nonlinear state therefore approaches the zero-energy regime as the support expands.

\begin{figure}[t]
\centering
\includegraphics[width=\linewidth]{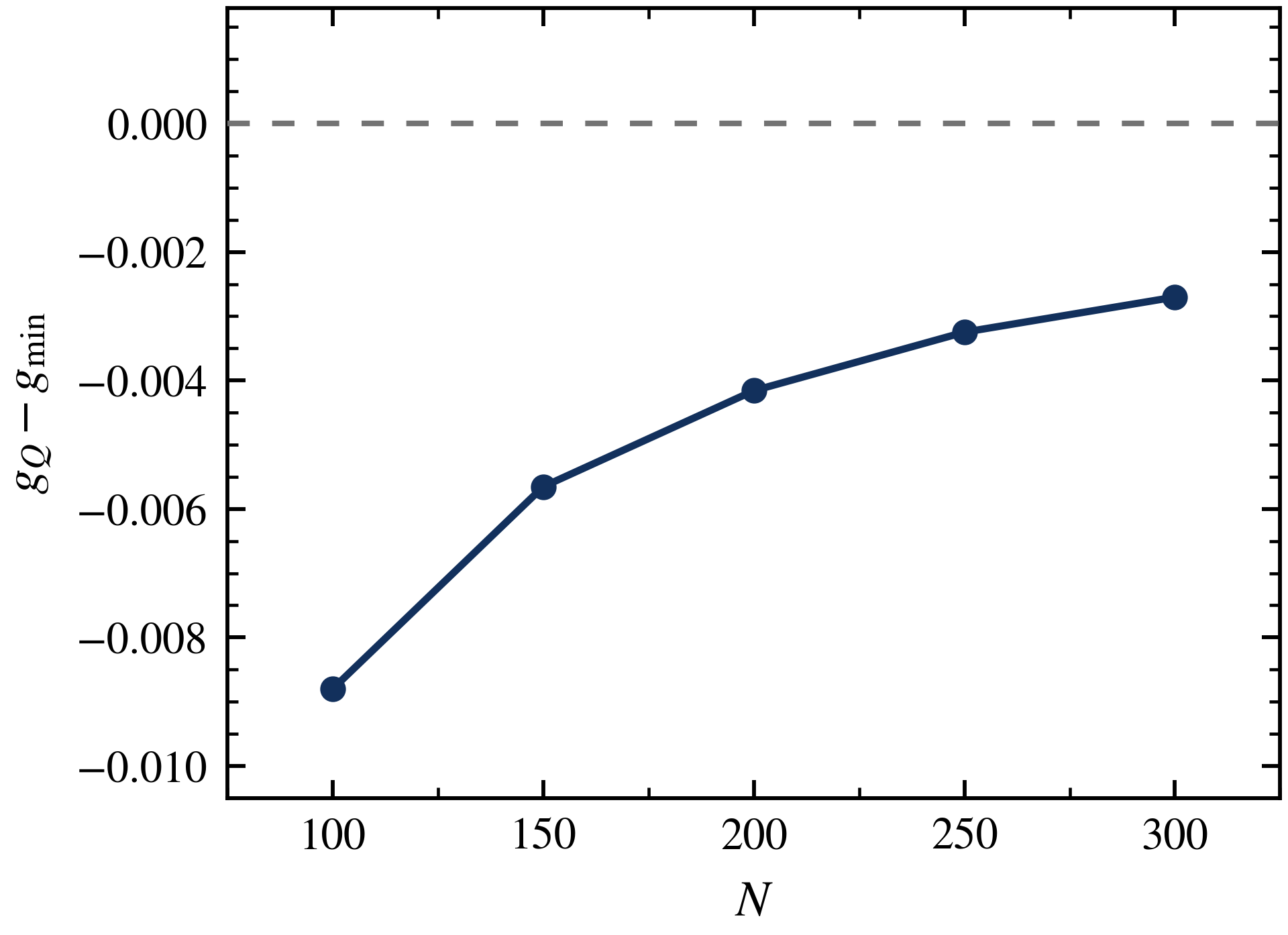}
\caption{Periodic case: Finite-size separation between the topological transition $g_Q$ and the intensity minimum $g_{\min}$ for $J_1 = 2$ and $J_2 = 1.7$. The difference $g_Q - g_{\min}$ remains negative and approaches zero with increasing $N$, showing that the two nonlinear scales become asymptotically coincident.}
\label{fig:FigA2}
\end{figure}

\section{Finite-size separation of $g_Q$ and $g_{\min}$}
\label{sec:finite-size-gQ-gmin}

In a periodic system, the topological transition and the intensity minimum occur at slightly different nonlinear strengths for a finite chain. We determine $g_{\min}$ from the minimum of $I^2$ and $g_Q$ from the condition
\begin{equation}
G_2 = G_1.
\label{eq:G2_G1_appendix}
\end{equation}

Figure~\ref{fig:FigA2} shows the difference $g_Q - g_{\min}$ for increasing system size. Its magnitude decreases with increasing $N$ and approaches zero. This shows that the two nonlinear scales become asymptotically coincident in the periodic system.

\begin{figure}[b]
\centering
\includegraphics[width=\linewidth]{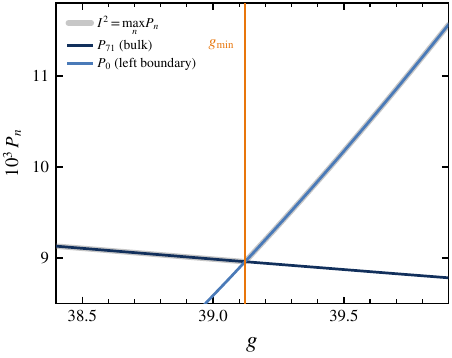}
\caption{Weak quasiperiodic case: Origin of the intensity minimum for $s = 0.5$, $J_1 = 2$, $J_2 = 1.7$, and $N = 144$. Below $g_{\min}$, the intensity $I^2$ follows the bulk probability $P_{71}$, while above $g_{\min}$ it follows the left boundary probability $P_0$. The vertical line marks $g_{\min}$.}
\label{fig:App_Fig3}
\end{figure}

\section{Origin of the weak-QP intensity minimum}
\label{app:weakmin}
In the weak quasiperiodic regime, the intensity minimum has a different origin from the periodic case. Figure~\ref{fig:App_Fig3} shows the probabilities of the cells that determine $I^2 = \max_n P_n$ near $g_{\min}$. Below the minimum, $I^2$ follows the bulk probability $P_{71}$. As $g$ increases, the probability at the left boundary grows and becomes the maximum. The two probabilities cross near $g_{\min} \simeq 39.12$. The intensity minimum therefore marks a change in the cell determining the maximum probability, from the bulk cell $n = 71$ to the left boundary $n = 0$.

\section{More discussions on threshold  nonlinearity strength to induce a topological phase transition}
\label{app:separation}

In the strong quasiperiodic regime the intensity minimum and the topological transition are governed by two different distances. The initial state is localized on a bulk cell; $g_{\min}$ is reached once probability has spread to the nearer boundary, a distance $d$ away, while $g_Q$ requires the geometric-mean condition $G_2=G_1$ to be met over all bonds and so is reached only once the state has spread throughout the whole chain, determined by the distance to the farther boundary, $D= N-1-d$. Because $d+D = N-1$, the two distances are complementary. A state near the boundary will have a small $g_{\min}$ and a large $g_Q$ while a centered state will likely have both values being close. In what follows we vary each distance while holding the other fixed. 

To vary $d$ at fixed chain length we use the phase of the quasiperiodic modulation. The intracell hopping is 
\begin{equation}
    t_{1,n} = J_1 + s\cos(2\pi\beta n + \varphi)\label{eq:phase}
\end{equation}
with $\varphi = 0$ throughout the main text. The phase does not change the geometric mean $G_1$ in the thermodynamic limit but at finite $N$ it selects which cells carry the weakest intracell bonds, and hence where the lowest positive energy eigenstate is localized. Shifting $\varphi$ therefore moves the localization center within a chain of fixed length. 
\begin{table}[b]
\caption{Route 1: the initial state is held on cell $26$, so $d = 26$ is
fixed while $D$ grows with the chain. $J_1 = 2$, $J_2 = 1.7$, $s = 5$,
$\varphi = 0$. Because the modulation of Eq.~\eqref{eq:phase} does not
depend on $N$, this is the same state in every row, with
$E = 1.13\times10^{-2}$ at $g = 0$ throughout; it is the lowest positive
energy eigenstate for $N \le 144$ and lies one and two levels higher at
$N = 180$ and $233$, where the longer chains carry weaker intracell bonds
at cells $170$ and $205$. The intensity minimum is unchanged across the
whole range while the topological transition grows in proportion to $D$.}
\label{tab:route1}
\begin{ruledtabular}
\begin{tabular}{ccccc}
$N$ & $d$ & $D$ & $g_{\min}$ & $g_Q$ \\
\colrule
 89 & 26 &  62 & 66.68 &  92.35 \\
110 & 26 &  83 & 66.68 & 133.73 \\
144 & 26 & 117 & 66.68 & 194.74 \\
180 & 26 & 153 & 66.68 & 252.17 \\
233 & 26 & 206 & 66.68 & 336.21 \\
\end{tabular}
\end{ruledtabular}
\end{table}
We first fix the localization center relative to the nearer boundary and lengthen the chain, so $d$ is held constant while $D$ grows. Table~\ref{tab:route1} lists the result for $J_1 =2$, $J_2 =1.7$, $s=5$ and $d=26$. The intensity minimum is unchanged over the whole range of chain lengths, while the topological transition grows in proportion to $D$. 

\begin{figure*}[t]
    \centering
    \includegraphics[width=\textwidth]{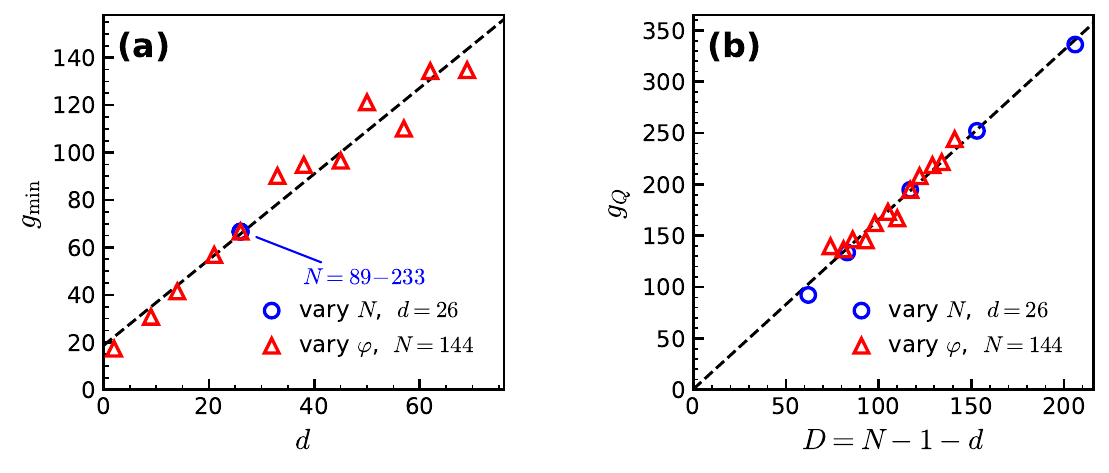}
    \caption{The two distances that set the nonlinear scales, at $J_1 =2$, $J_2 =1.7$ and $s = 5$. (a)~$g_{\min}$ against the distance $d$ to the nearer boundary. (b)~$g_Q$ against the distance $D = N-1-d$ to the farther boundary. Circles: the chain is lengthened from $N=89$ to $N=233$ with the initial positive energy state being localized at cell $26$, so $d$ is constant while $D$ grows. Triangles: the chain is fixed at $N=144$ and the modulation phase $\varphi$ is varied, which relocates the localization center, so $d$ and $D$ change while the chain length is constant. The five circles in (a) are superimposed, $g_{\min}$ is unchanged across the whole range of chain lengths, and the dashed line there is fitted to the phase data alone. In (b) the two sets fall on a common line, so $g_Q$ does not distinguish cells added by lengthening the chain from cells acquired by moving the state.} 
    \label{fig:appD_nearfar}
\end{figure*}

We then hold the chain length fixed at $N=144$ and shift $\varphi$ through one period, which moves the localization center across the chain. Here $g_{\min}$ and $g_Q$ move in opposite directions. As the initial state is placed further away from the nearer boundary, $g_{\min}$ increases while $g_Q$ decreases, since here $d$ increases while $D$ decreases. The separation is the largest when the state begins close to a boundary and smallest when it begins near the center. Figure~\ref{fig:appD_nearfar} shows both sets of data plotted against the relevant distances. Although the two procedures change $D$ by entirely different means, adding lattice sites in one case and relocating the state in the other, the resulting values of $g_Q$ fall on a common line, and likewise for $g_{\min}$ against $d$. 

\begin{figure*}[t]
    \centering
    \includegraphics[width=\textwidth]{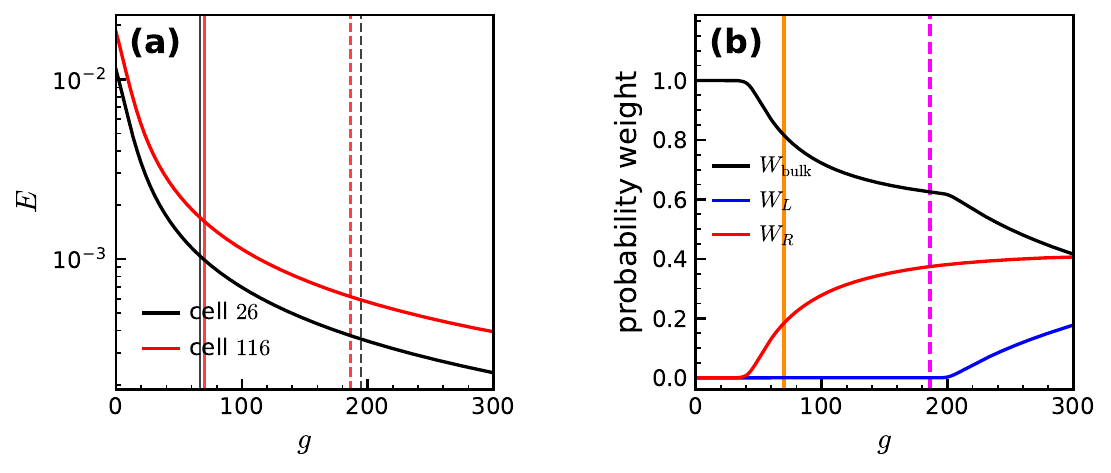}
    \caption{Two different initial states at $J_1=2$, $J_2=1.7$, $s=5$ and $N=144$. (a)~Eigenenergy $E$ as a function of the nonlinear strength for the lowest positive energy state, localized on cell $26$, and for a higher energy state localized on cell $116$, with $E(g=0) = 1.13 \times 10^{-2}$ and $1.83 \times 10^{-2}$ respectively. Both decay monotonically and show no feature at either $g_{\min}$ or $g_Q$. (b)~Probability weights $W_{\mathrm{bulk}}$, $W_L$, and $W_R$ for the higher energy state as defined in Fig.~\ref{fig:appD}. In (a), the vertical lines follow the color of the corresponding curve, solid for $g_{\min}$ and dashed for $g_Q$; in (b), the orange solid and magenta dashed lines mark $g_{\min}$ and $g_Q$, respectively.}
    \label{fig:appD_seed}
\end{figure*}

The same behavior can be observed when the initial state is not the lowest positive energy eigenstate. For $N=144$, we repeat the continuation from a higher energy eigenstate localized on cell $116$ which lies $27$ cells from the right edge and so has essentially the same $d$ as the lowest-energy state used in the main text, localized on cell $26$. The two give $g_{\min} = 70.19$ and $66.68$, and $g_Q=186.15$ and $194.74$ respectively. Figure~\ref{fig:appD_seed} shows the eigenenergy and the probability weights along this branch.

\begin{figure*}[t]
    \centering
    \includegraphics[width=\textwidth]{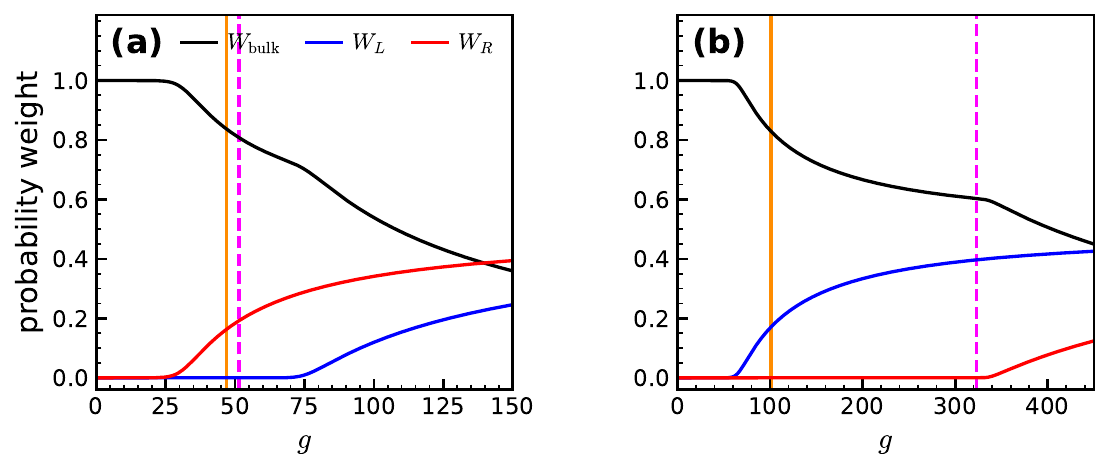}
    \caption{Probability weights $W_{\mathrm{bulk}}$, $W_L$, and $W_R$ as functions of the nonlinear strength, the boundary weights are defined over the first and last ten cells, with the remaining cells defining the bulk, as in Fig.~\ref{fig:Fig10_Strong_QP}(b). (a)~$s=4.1$ and (b)~$s = 6$, with $J_1 = 2$, $J_2 = 1.7$ and $N=144$. The initial state is localized on cell $95$, $48$ cells from the right edge for (a) and it is localized on cell $27$, $27$ cells from the left edge for (b) respectively. Note the different horizontal ranges of the two panels. The orange solid and magenta dashed lines mark $g_{\min}$ and $g_Q$, respectively.}
    \label{fig:appD_s}
\end{figure*}

The same structure holds elsewhere in the strong quasiperiodic regime. We chose the values $s=4.1$ and $s=6$ and use the lowest positive energy eigenstate as the initial state. As shown in Fig.~\ref{fig:appD_s}, for $s=4.1$, the initial state is localized at cell $95$, making it $48$ cells away from the right edge, so $D$ and $d$ are closer, resulting in $g_{\min}$ and $g_Q$ being closer together, having values $46.93$ and $51.41$ respectively. For $s=6$, similar to $s=5$, it is localized near the left edge, hence having a larger separation, with  $g_{\min} = 101.15$ and $g_Q = 322.78$.

\begin{figure*}[t]
    \centering
    \includegraphics[width=\textwidth]{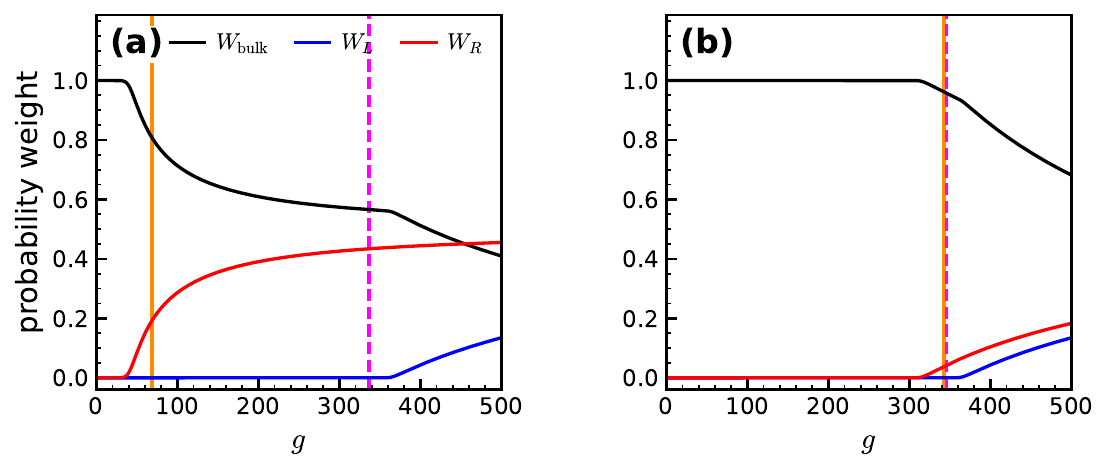}
    \caption{Probability weights $W_{\mathrm{bulk}}$, $W_L$, and $W_R$ as functions of the nonlinear strength, the boundary weights are defined over the first and last ten cells, with the remaining cells defining the bulk, as in Fig.~\ref{fig:Fig10_Strong_QP}(b). (a)~$N=233$ and (b)~$N = 377$, with $J_1 = 2$, $J_2 = 1.7$ and $s=5$. The initial state is localized on cell $205$ in both. The chain length places this $27$ cells from the right edge at $N=233$, and close to the center of the chain for $N=377$, giving $g_{\min} = 68.60$ and $342.51$ respectively. The orange solid and magenta dashed lines mark $g_{\min}$ and $g_Q$, respectively.}
    \label{fig:appD}
\end{figure*}

When the initial state happens to lie near the middle of the chain, the two distances nearly coincide and the separation almost vanishes. At $N=377$ the lowest positive energy eigenstate is localized near the center of the chain, and $g_{\min}$ and $g_Q$ differ by about one percent. Figure~\ref{fig:appD} shows the corresponding probability weights. 

\bibliography{apssamp}

\end{document}